\documentclass[sigconf]{acmart}
\usepackage{amsmath}
\usepackage{graphicx}
\usepackage{ragged2e}

\AtBeginDocument{%
  }

\setcopyright{acmlicensed}
\copyrightyear{2018}
\acmYear{2018}
\acmDOI{XXXXXXX.XXXXXXX}
\acmConference[Conference acronym 'XX]{Make sure to enter the correct
  conference title from your rights confirmation email}{June 03--05,
  2018}{Woodstock, NY}
\acmISBN{978-1-4503-XXXX-X/2018/06}

\begin{document}

\title{Addressing Information Loss and Interaction Collapse: A Dual Enhanced Attention Framework for Feature Interaction}

\author{Yi Xu}
\affiliation{%
  \institution{Lazada Group}
  \city{Beijing}
  \country{China}}
\email{xy397404@alibaba-inc.com}

\author{Zhiyuan Lu}
\affiliation{%
  \institution{Beijing University of Posts and Telecommunications}
  \city{Beijing}
  \country{China}}
\email{luzy@bupt.edu.cn}

\author{Xiaochen Li}
\affiliation{%
  \institution{Lazada Group}
  \city{Beijing}
  \country{China}}
\email{xingke.lxc@lazada.com}

\author{Jinxin Hu}

\authornotemark[1]
\affiliation{%
  \institution{Lazada Group}
  \city{Beijing}
  \country{China}}
\email{jinxin.hjx@lazada.com}

\author{Hong Wen}
\affiliation{%
  \institution{Unaffiliated}
  \city{Beijing}
  \country{China}}
\email{dreamonewh@gmail.com}

\author{Zulong Chen}
\affiliation{%
  \institution{Alibaba Group}
  \city{Beijing}
  \country{China}}
\email{chenzulong198867@gmail.com}

\author{Yu Zhang}
\affiliation{%
  \institution{Lazada Group}
  \city{Beijing}
  \country{China}}
\email{daoji@lazada.com}

\author{Jing Zhang}
\affiliation{%
  \institution{Wuhan University, School of Computer Science	}
  \city{Wuhan}
  \country{China}}
\email{jingzhang.cv@gmail.com}

\authornote{Corresponding Author.}
\begin{abstract}
The Transformer has proven to be a significant approach in feature interaction for CTR prediction, achieving considerable success in previous works. However, it also presents potential challenges in handling feature interactions. Firstly, Transformers may encounter information loss when capturing feature interactions. By relying on inner products to represent pairwise relationships, they compress raw interaction information, which can result in a degradation of fidelity. Secondly,  due to the long-tail features distribution, feature fields with low information-abundance embeddings constrain the information abundance of other fields, leading to collapsed embedding matrices. To tackle these issues, we propose a Dual Attention Framework for Enhanced Feature Interaction, known as Dual Enhanced Attention. This framework integrates two attention mechanisms: the Combo-ID attention mechanism and the collapse-avoiding attention mechanism. The Combo-ID attention mechanism directly retains feature interaction pairs to mitigate information loss, while the collapse-avoiding attention mechanism adaptively filters out low information-abundance interaction pairs to prevent interaction collapse. Extensive experiments conducted on industrial datasets have shown the effectiveness of Dual Enhanced Attention.
\end{abstract}



\maketitle

\section{Introduction}

The Transformer architecture~\cite{attention} has served as the foundational enabler for recent breakthroughs in large language models (LLMs), which have revolutionized fields such as natural language processing\cite{llama}, multimodal\cite{qwen}, etc. In click-through rate (CTR) prediction, Transformer has been successful with its powerful context-aware capabilities~\cite{AutoInt,bert4rec,sasrec} and has inspired a series of work on context-aware recommedations\cite{context,masknet}. Its paradigm integrates traditional feature-crossing methods such as factorization machines (FM)\cite{FM,NFM,afm} while introducing global contextual awareness through softmax-based attention mechanisms. 
However, despite its demonstrated success, we identify two critical limitations of Transformers in feature interaction: a) Imformation loss of feature interaction and b) Interaction collapse. 
\begin{figure}
    \centering
    \includegraphics[width=0.8\linewidth]{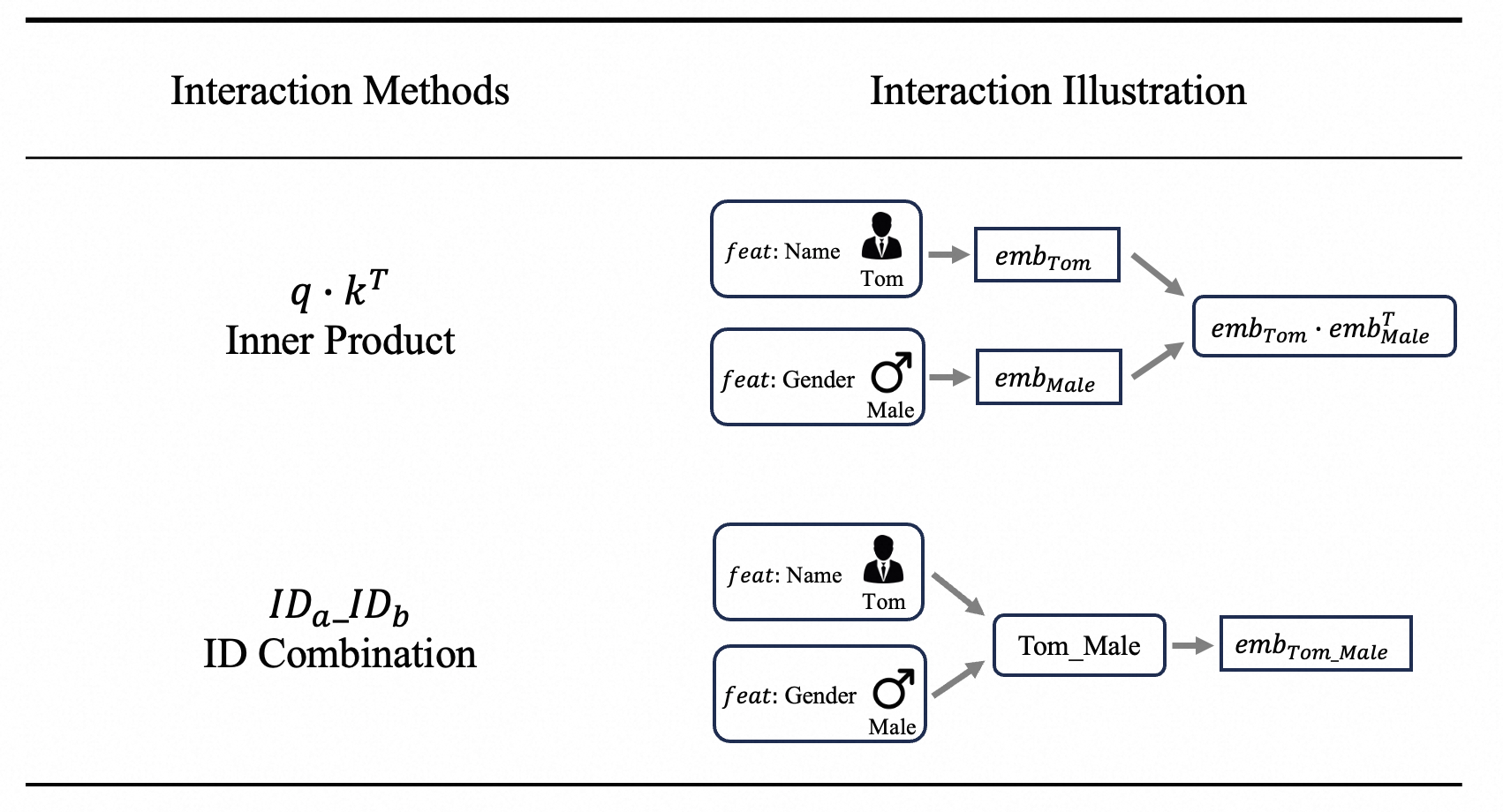}
    \caption{Comparision of Inner-product and Combo-ID}
    \label{fig:enter-label}
    \vspace{-18pt}
\end{figure}
\textbf{Imformation loss of feature interaction} The feature interaction methods such as inner product\cite{FM,NFM,afm,DeepFM}, out product\cite{DCNv2,xdeepfm}, or bilinear feature interactions\cite{Fibinet,Zhang2022FiBiNetRM,FinalMLP} cannot express the empirical feature interactions precisely, where is information loss due to incomplete representation capacity. For example, the embeddings of "Apple" and "Orange" are close in representation space, which easily leads to that the inner products of (Steve Jobs, Apple) and (Steve Jobs, Orange) are also close. But in fact, there is little relationship between Steve Jobs with Orange, leading to suboptimal model performance. A direct solution is Combo-ID, which means that combine feature interaction pair to a new ID and assigns embeddings for feature interaction pairs, which learns feature interaction more precisly.Previous works such as CAN\cite{CAN} and MemoNet\cite{Memonet} employ methods similar to Combo-ID for feature interaction. However, due to the combinatorial explosion of feature interactions and the constraints of online storage, the hash collisions lead to the confused representations.

\textbf{Interaction collapse:} Recommendation systems fundamentally differ from LLMs as they handle  the continuous changing billion-scale features vocabularies, which refers to user IDs, item IDs, and merchant attributes. 
 Long-tailed features are prone to obtain embedding matrices that were not trained enough, which can limit the information abundance of other feature fields and lead to the interaction collapse problem of feature interaction, as mentioned in \cite{MultiEMb}.
However, the challenge of addressing interaction collapse within transformer-based models remains underexplored.

To address these challenges, we propose a dual
attention framework to enhanced feature interaction  efficiently(Dual Enhanced Attention) in CTR prediction. Specifically, we provide insightful solutions from 2 perspectives:

\textbf{Combo-ID Attention Mechanism} is proposed to alleviate the information loss of attention mechanism on feature interaction, we introduce an independent memory mechanism that allocates learnable representations to each feature interaction pair, and recomputes the attentions scores to enhance feature interaction. Furthermore, to mitigate the problem of confuse information representation caused by hash collisions, we proposed the gated simaese codebook.

\textbf{Collapse-avoiding Attention Mechanism} is responsible for the generalization of feature interaction, and in order to avoid the interaction collapse caused by long-tailed features, we adaptively select the top feature interactions.



In summary, Our key contributions are summarized as follows:
\begin{itemize}
    \item To alleviate the
    information loss of attention mechanism on feature interaction, we propose the Combo-ID attention mechanism to enhance the representation ability of feature interactions.
    \item To enhance the  generalization of the attention Mechanism in feature interaction, we propose the collapse-avoiding attention mechanism.
    \item We evaluate our proposed method on industrial datesets, demonstrating its effectiveness through extensive experiments.
\end{itemize}

\section{PRELIMINARY}
In this section, we define the overrall workflow when Transformer is applied to CTR prediction task.

\textbf{Inputs Layer} Recommendation models are trained with a large amout of features from multiple perspetives, can be categoried into sparse features and dense features. Dense features are discretize to be assiged IDs with general bucketing strategy. The input features can be formulated as: $x=[ID_{f_1},ID_{f_2},\dots ,ID_{f_n}]$, where $n$ denotes the number of feature fields and $f_i$ denotes the $i$-th feature field.
Each feature field has a hash table for storing its embeddings. For each feature field, the feature IDs are mapped to different addresses of the embedding matrix by a general hash function. The input layer can be formulated as:
\begin{equation}
    Q=K=V=[e_1,e_2,\dots ,e_n]
\end{equation}
where $\boldsymbol{e_i}\in\mathbb{R}^{d}$ denotes the embedding of one field and $d$ is the embedding dimension. Each feature can be regarded as a token, and in attention based recommendation model, Q, K and V are the sequence of input features equivalently. 

\textbf{Attention-based Feature Interaction Module}
When apply transformer for modeling feature interaction, each self-attention layer captures 2-order feature interaction relationships for the tokens output by the last layer. Each element of the attention weight matrix is a representation of a feature interaction pair. Futhermore, flatten all tokens and feed them into a DNN for CTR prediction. The formulation is as follows:
\begin{align}
    V^{\prime}=Attention(Q,K,V) \\
    y=\delta(DNN(flatten(V^{\prime}))
\end{align}

\section{Methodology}
In this section, we introduce the proposed method which consists of 3 parts, the Combo-ID attention mechanism, collapse-avoiding attention mechanism, and fusion mechanism. 
\begin{figure*}[h]
    \centering
    \includegraphics[width=0.6\textwidth]{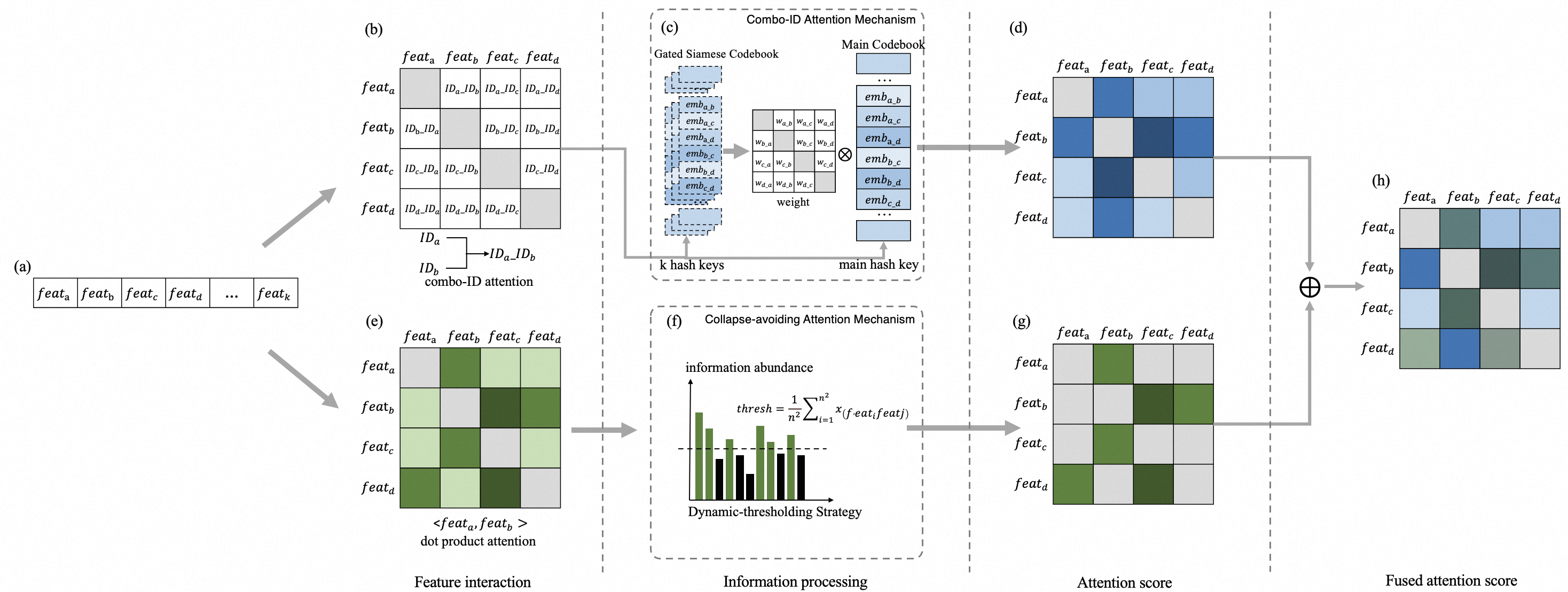}
    \captionsetup{justification=justified} 
    \caption{\justifying Illustration of Combo-ID Attention Mechanism and Collapse-avoiding Attention Mechanism: (a) A set of input feature, e.g. user IDs, item IDs, and merchant attributes. (b) In the Combo-ID attention mechanism, each pair of features is combined by generating a unique Combo-ID through concatenation of their individual feature IDs. (c) The Gated Siamese Codebook method employs $k$ codebooks with distinct hash functions, using siamese representations to gate and re-weight the main codebook's outputs, reducing misrepresentation of long-tail feature interactions. (d) After re-weighting the main codebook, each embedding is projected into a scalar, eventually forming the attention score matrix of the Combo-ID Attention. (e) The traditional self-attention uses inner product to calculate attention score. (f) The dynamic-thresholding strategy filters out low information abundance embeddings by using the average modulus length within a batch as a threshold. (g) The attention score matrix of the Collapse-avoiding Attention Mechanism. (h) The final attention score matrix is fused by the Combo-ID Attention Mechanism and the Collapse-avoiding Mechanism Attention.}
    \label{fig:model_detail}
\end{figure*}
\subsection{Combo-ID Attention Mechanism}
The Combo-ID attention mechanism module memorizes all feature interactions automatically with an independent memory mechanism. Codebook is a storage location for attention knowledge, where each codeword is a minimal storage unit for storing the representation of a feature interaction pair. 

\textbf{Combo-ID Memorization.} Define codebook $\boldsymbol{C}$ as parameter matric $\boldsymbol{C}\in\mathbb{R}^{s\times d}$, the size of codebook is ${s}$ and dimension is ${d}$. Each row of parameter matric $\boldsymbol{C}$ is a codeword whcih is a vector with dimension of ${d}$.

For a feature interaction pair $(feat_i,feat_j)$ which means the combination of feature $feat_i$ and feature $feat_j$. Semantically, the ID of $(feat_i,feat_j)$ comes from the features $feat_i$ and feature $feat_j$. 
\begin{equation}
    ID_{(feat_i,feat_j)}=[ID_{feat_i},ID_{feat_j}]
\end{equation}
Through a general hash function projects, we get the address $a_{(i,j)}$ of the feature cross pair $(feat_i,feat_j)$ in Codebook as follows:
\begin{equation}
    a_{(i,j)}=H(ID_{(feat_i,feat_j)})
\end{equation}
where $H$ is the hash function. 

\textbf{Gated Siamese Codebook.}
Due to the combinatorial explosion problem of feature interactions, compressing all features interaction pairs into a fixed-size codebook by hashing is storage-efficient and can be served online. With constrained codebook size, codebooks have hash collisions that cause each codeword unavoidably mixes representations of different feature interactions. According to whether the features have enough samples to be well-trained or not, they can be categorized into well-trained features and under trained features. Depending on the type of collision, the impact is shown in Fig.\ref{fig:enter-label}. 
\begin{figure}
    \centering
    \includegraphics[width=0.8\linewidth]{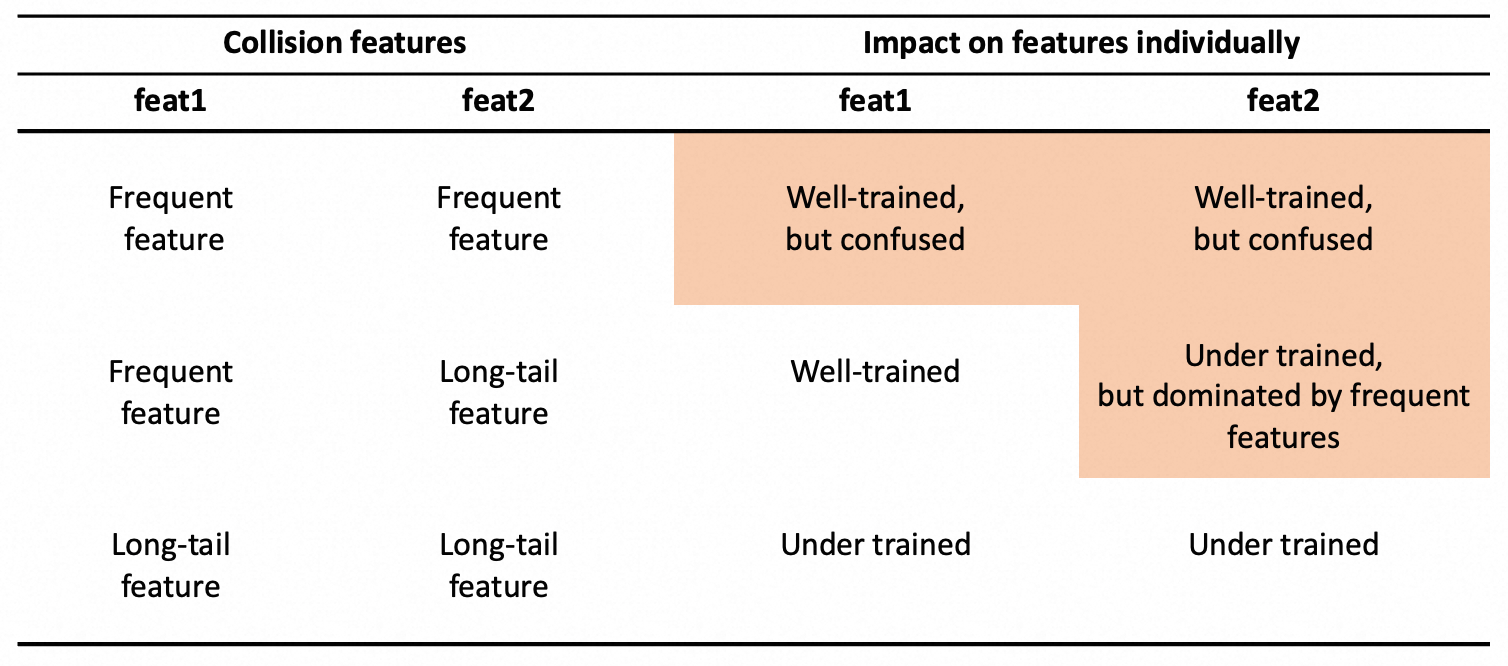}
    \caption{The impacts of collision on features individually}
    \label{fig:enter-label}
    \vspace{-18pt}
\end{figure}
Among them, the problem between well trained, between well-trained and under trained is the most serious. to address this problem, we design a way to reduce the impact of collision by k siamase hashtables. Assuming that the collision probability of two features in a codebook is $p$, $k$ hashtable have different hash key and are independently, so the collision probability in $k$ hashtable is $p^k$, since $p<1$. Theoretically, the larger $k$, the lower the collision probability, and we utilize the representations of siamase codebooks to vote for the main codebook.
For each feature interaction pair $(feat_i,feat_j)$,the representation $e_{(i,j)}$ can be formulated as follows:
\begin{equation}
e_{(i,j)}= \phi(W_0[C^1_{a^1_{(i,j)}},C^2_{a^2_{(i,j)}},\dots,C^k_{a^k_{(i,j)}}])\cdot C_{a_{(i,j)}}
\end{equation}
where $C$ is the main codebook, $C^1, C^2,\dots,C^n$ are siamese codebooks which has the same size with $C$, $C^1_{a^1_{(i,j)}}$ is the representation of feature interaction pair $(feat_i,feat_j)$ in codebook $C^1$, $\boldsymbol{W_0}\in\mathbb{R}^{d_w\times 1}$, $d_w=kd$, and $\phi$ is non-linear activation. According to the representations of the simase codebooks , $\phi(W_0[C^1_{a^1_{(i,j)}},C^2_{a^2_{(i,j)}},\dots,C^k_{a^k_{(i,j)}}])$ is the score of the simase codebooks.

For each element in the attention matrix, we address the representation from the codebook and combine these representations to obtain the attention score matrix for each layer, which is characterized as 
$\boldsymbol{E}\in\mathbb{R}^{n\times n \times d}$. $n$ is the number of features. $d$ is the dimension of feature embedding. 
\begin{equation}
    E=\begin{bmatrix}
        e_{q_1,k_1},e_{q_1,k_2}\dots e_{q_1,k_n},\\
        e_{q_2,k_1},e_{q_2,k_2}\dots e_{q_2,k_n},\\
        \dots \\
        e_{q_n,k_1},e_{q_n,k_2}\dots e_{q_n,k_n},\\
    \end{bmatrix}
\end{equation}

\textbf{Attention Re-weight.} We leverage the representation of feature interactions $\boldsymbol{E}$ 
to calculate the attention matrix. A subnetwork is designed to project each feature interaction representation into an attention score, as follows:
\begin{gather}
     a_{i,j}=W_2(f_1(W_1(e_{q_i,k_j})+b_1))+b_2\\
      A=\begin{bmatrix}
        a_{1,1},a_{1,2}\dots a_{1,n},\\
        a_{2,1},a_{2,2}\dots a_{2,n},\\
        \dots \\
        a_{n,1},a_{n,2}\dots a_{n,n},\\
    \end{bmatrix}\\
    A_m=A-diag(A)
\end{gather}

where $\boldsymbol{W_1}\in\mathbb{R}^{d\times h}$, $h$ is the dimension of hidden layer, $\boldsymbol{W_1}\in\mathbb{R}^{h\times 1}$ are parameter matrix of two MLP layers, $f_1$ is the non-linear activation function. The parameters of this DNN are shared among all feature interaction pairs. Then, the attention matrix is reshaped into
$\boldsymbol{A}\in\mathbb{R}^{n\times n}$.
To avoid self-cross dominant training leading to suboptimal results, the diagonal matrix of the attention matrix is removed, the revised attention score matrix denotes $\boldsymbol{A_m}$.

\subsection{Collapse-avoiding Attention Mechanism}
The vanilla self-attention has universal generalizability on automatic modeling of feature interaction, but is prone to interaction-collapse problems on large-scale recommendation tasks. To boost the performance of attention in feature interaction and avoid interaction-collapse, the collapse-avoid strategy is employed to filter long-tail feature interactions. 

The phenomenon of interaction-collapse occurs when modeling the feature interaction between long-tail features and other features, the low-rank embedding of the long-tailed features constrains the information abundance of the other features, which leads to suboptimal performance of the model.


\textbf{Dynamic-thresholding Strategy}
Ideally, filtering out the low-information-abundance embedding is a direct solution to interaction-collapse. During the training process, the information abundance is difficult to compute, and the modulus length identifies the sparsity of the embedding matrix, the larger the modulus length, the lower its sparsity. Considering the long-tailed distribution of features and the dynamic updating of mode length during training, we compute the average of the mode length of the representation of feature interaction within the batch as the threshold. The formulation is as follows:
\begin{gather}
    A_c=Thresh(Q K^\mathsf{T}-diag(Q K^\mathsf{T}))\\
   Thresh(\cdot)=\begin{cases}
1 & \text{if } x \geq thresh \\
0 & \text{if } x < 0
\end{cases}\\
thresh= \frac{1}{n^2} \sum_{i=1}^{n^2} 
||x_{(fest_i,feat_j)}||_2\\
x_{(fest_i,feat_j)}=e_i\cdot e_j^\mathsf{T}
\end{gather}
where $thresh$ is the average modulus length of the feature embedding.
\subsection{Fusion Mechanism}
Considering the distributions of the combo-ID attention scores $A_m$ and collapse-avoiding attention scores $A_c$ is quietly different, we propose 3 types of fusion mechanisms, including weighted sum, gated balance and multiply. The formulation of weighted sum is as follows:
\begin{equation}
    Attention(Q,K,V)=softmax(\alpha\cdot A_m+\beta\cdot A_c)V
    \label{eq:fusion1}
\end{equation}
where $\alpha$ and $\beta$ are learnable parameters. To balance the generalization and memoraization better, the gated balance method and multiply-based method are also proposed and employed in experiments. Gated balance method assumes that the two attention mechanisms are complementary as shown in Eq.\ref{eq:fusion2}, and multiply-based treats collapse-avoiding as a revision of the Combo-ID weighting to enhance the effect of top feature interaction as shown in Eq.\ref{eq:fusion3}. The formulations is as follows: 
\begin{equation}
    Attention(Q,K,V)=softmax((1-g(A_c))\cdot A_m+g(A_c)\cdot A_c)V
    \label{eq:fusion2}
\end{equation}
\begin{equation}
    Attention(Q,K,V)=softmax(2\cdot \sigma(W_2A_c) A_m+g\cdot )V
    \label{eq:fusion3}
\end{equation}
where $g(\cdot)$ is a MLP layer with non-linear activation function.
\section{Experiment}
\subsection{Experiment Setup}
\textbf{Datasets:} To verify the effectiveness of the proposed method, we have conducted experiments on industrial datasets collected from
the advertising systems of a leading Southeast Asian e-commerce
platform. The industrial dataset contains 500 million records.

\textbf{Models for Comparision:} We compare the proposed method with FM\cite{2010FM},AFM\cite{afm},AutoInt\cite{AutoInt},Fibinet\cite{Fibinet},MemoNet\cite{Memonet},HSTU\cite{HSTU}. FM, AFM,AutoInt, Fibinet, and MemoNet are representative works on feature interaction at different periods of time, where AFM, AutoInt are attention based methods. MemoNet has proposed to memorize all possible cross features with a multi-hash codebook to enhance the memorization of CTR models. HSTU has proposed a new encoder designed for feature interaction and demonstrated scaling laws of a new deep learning recommendation model formulation. 

\textbf{Evaluation Metrics:} For the evaluation, we use the widely used AUC and GAUC\cite{GAUC} as previous works, and GAUC is the most important metric for our personalized ads system.
\begin{equation}
GAUC = \sum_{s} w_s AUC_s \quad \text{where} \quad w_{session} = \frac{\#logs_{session}}{\sum_{i} \#logs_i}
\end{equation}
where the $w_{session}$ denotes the logs ratio of the session.
\subsection{Performance Comparision}
\textbf{Overall Performance.} We report the performance of baselines and the proposed method in Table.\ref{tab:Performance}. The proposed method outperforms the baselines. Fibinet introduces bilinear feature interaction to improve the expressive ability of feature interactions, and Memonet improve the memorization ability of the CTR models by memorizing the key interactions through an independent memorization mechanism. The results of both Fibinet and MemoNet exceed AutoInt, which verifies that there are problems of lack of memorability and expressive ability in attention mechanism. The GAUC improvement of this proposed method compared with AutoInt can demonstrate the effectiveness of Dual Enhanced Attention.
\vspace{-8pt}
\begin{table}[htbp] 
  \centering  
  \begin{tabular}{ccc}  
    \toprule
    Model & AUC & GAUC \\  
    \midrule
    FM & 67.54 & 60.00 \\
    Fibinet & 68.29 & 60.42\\
    MemoNet & 68.17 & 60.41\\
     \midrule
    AFM & 67.14 & 60.12\\
    AutoInt & 68.20 & 60.34  \\  
    HSTU & 68.19 & 60.41  \\  
    \textbf{Dual Enhanced Attention} & \textbf{68.32}  & \textbf{60.47}  \\  
    \bottomrule
  \end{tabular}
  \caption{Performance Comparision on Industrial Dataset}  
  \label{tab:Performance}  
  \vspace{-18pt}
\end{table}
\subsection{Ablation Analysis}
 \textbf{Effectiveness of Each Component.}Several ablation studies have been conducted to investigate the effectiveness of each component, as shown in Table \ref{tab:Ablation}.Firstly, on the recommendation task, the diagonal matrix of attentions represents the self-crossing of features, which may lead to suboptimal learning of the model, and the improvement of “transformer w/o diag” supports this hypothesis. Further, Collapse-avoiding Attention Mechanism removes the long-tailed feature interaction based on “transformer w/o diag”, which brings further improvement. In addition, we individually validate the effect of Combo-ID Attention Mechanism, which brings about an increase in GAUC, proving that the introducing an independent memory mechanism to memorize feature interactions is effective. 
\begin{table}[tbp] 
  \centering  
  \begin{tabular}{ccc}  
    \toprule
    Model & AUC & GAUC \\  
    \midrule
    Transformer & 68.18 & 60.31  \\  
    Transformer(w/o diag)  & 68.22(+0.04pt) & 60.34(+0.03pt) \\  
    Combo-ID Attention & 68.05(-0.13pt) & 60.40(+0.09pt) \\
    Collapse-Avoiding Attention & 68.19(+0.01pt) & 60.43 (+0.12pt) \\
    \textbf{Dual Enhanced Attention} & 68.32\textbf{(+0.14pt)} & 60.47\textbf{(+0.16pt)} \\  
    \bottomrule
  \end{tabular}
  \caption{Ablation Studies on Industrial Dataset}  
  \label{tab:Ablation}  
\end{table}

\textbf{Analysis of Combo-ID Attention Mechanism}
In Combo-ID Attention Mechanism, we point the hash collision problem of memorizing feature interactions and accordingly propose the \textbf{g}ated \textbf{s}iames \textbf{c}odebook(gsc) to solve these problems. As shown in Table.\ref{tab:gsc}, the results of the ablation experiments has demonstrated the effectiveness of gsc.

\begin{table}[htbp] 
  \centering  
  \begin{tabular}{ccc}  
    \toprule
    Model & AUC & GAUC \\  
    \midrule
    Combo-ID Attention(w/o gsc) & 68.05 & 60.40 \\
    Combo-ID Attention & 68.12(+0.07) & 60.45(+0.05pt) \\
    \bottomrule
  \end{tabular}
  \caption{Analysis of Combo-ID Attention Mechanism}  
  \label{tab:gsc}  
\end{table}

\textbf{Analysis of Fusion Mechanism}
For the balance the dual attention mechanisms, we conduct experiments with three fusion methods, weighted sum, gated Balance and multiply.  As shown in Table.\ref{tab:Fusion}, there are a small gaps between 3 fusion mechanism.
\begin{table}[htbp] 
  \centering  
  \begin{tabular}{ccc}  
    \toprule
    Model & AUC & GAUC \\  
    \midrule
    Weighted Sum & 68.29 & 60.48  \\  
    Gated Balance & 68.32 & 60.44 \\  
    Multiply & 68.32 & 60.47 \\
    \bottomrule
  \end{tabular}
  \caption{Analysis of the variants of Fusion Mechanism}  
  \label{tab:Fusion}  
\end{table}
\vspace{-15pt}
\section{Conlusions}
In this paper, we analyze the problem of information loss and interaction collapse of Transformer applied to large-scale feature interaction. Accordingly, we propose a dual enhanced attention framework for feature interaction in CTR prediction and conduct Extensive experiments on industrial dataset to demonstrate the effectiveness.
\bibliographystyle{ACM-Reference-Format}
\bibliography{main}



\end{document}